# Measuring $^{241}$Am Dipole Response

(Primakoff Photon-Magnetic Field Coupling Experiment)

C. Scarlett[a], E. Fischbach[b], B. Freeman[b], J. J. Coy[c], P. Edwards[d], D. Osborne[a], J. Edwards[a], L. Mwibanda[b], and A. Alsayegh[b]

[a] Department of Physics, Florida A&M University

[b] Department of Physics and Astronomy, Purdue University

[c] Department of Computer Science, Ball State University

[d] Department of Physics, College of Coastal Georgia

## ABSTRACT:

Americium ($^{241}$Am) with an unpaired proton in the $F_{5/2}$ state exhibits a significant magnetic dipole moment. The dipole can be experimentally measured with application of even modest external magnetic fields, as little as 1G, as a shifting in the energy spectrum of emitted gammas during the process of decaying to $^{237}$Np ground state. This paper looks at the shifting in the output energy peak of gammas from the decay of excited $^{237}$Np when two configurations of an external magnetic field are applied. The peak shifting, which does not appear in the background data dominated by $^{238}$U decays, differs for the two dominant gammas released at 26.3 keV and 59.5 keV. For the 59.5 keV peak: shifting is ~ 32% of 1-Energy Bin or about 0.5 keV. While for the 26.3 keV peak: shifting is ~ 15% of 1-Energy Bin or about 0.24 keV. Interestingly enough, there appears to be a shifting for the case where the field remains in a direction horizontal to the optical bench and the light is simply blocked or unblocked from entering the field, referred to as the light (sP) or dark (sD) modes.

## Introduction:

The absence of strong CP violations, given that such effects have been measured for weak interactions, has long puzzled particle physicists. Adding to the mystery is the apparent lack of a measurable neutron electric dipole moment (nEDM), which should arise from the quark content of the nucleons. An eloquent solution to both of these observations was proposed by Peccei-Quinn[1] as the coupling between a scalar field and at least one fermionic particle. This scalar field also gives rise to a boson particle dubbed the "axion." A number of experiments have been proposed and executed to search for this new particle, which in theory should result from interactions between photons and external magnetic fields. The photon-magnetic field interaction that could directly produce axions had already been theorized by Henry Primakoff as far back as 1951 to explain photo-production of mesons in the Coulomb field of a nucleus[2].

The presence of a magnetic field in space breaks the isotropy of space. Combined with the Peccei-Quinn scalar field, this broken isotropy should lead to absorption and/or polarization rotation whenever a photon beam passing through a magnetic field. Experimental searches for axion production and decay have largely focused on finding the lowest order Feynman diagram whereby photons beams rotate or disappear leading to optical effects that could be measured (see Figure 1 Left). After numerous experimental searches (PVLAS, BNL E840, BMV, etc.)[3-5] no group has yet seen an effect that demonstrates the production of axions.

The experiment described here takes a look at a higher order effect (see Figure 1 Right) where the production of axions and/or weakly interacting particles is just part of the equation. The current approach aims those weak particles at a radioactive sample to search for changes in the detected number of decays. Such a process involves two Feynman diagrams, one for the production of weak particles and the other for the absorption of these particles by a nucleus.

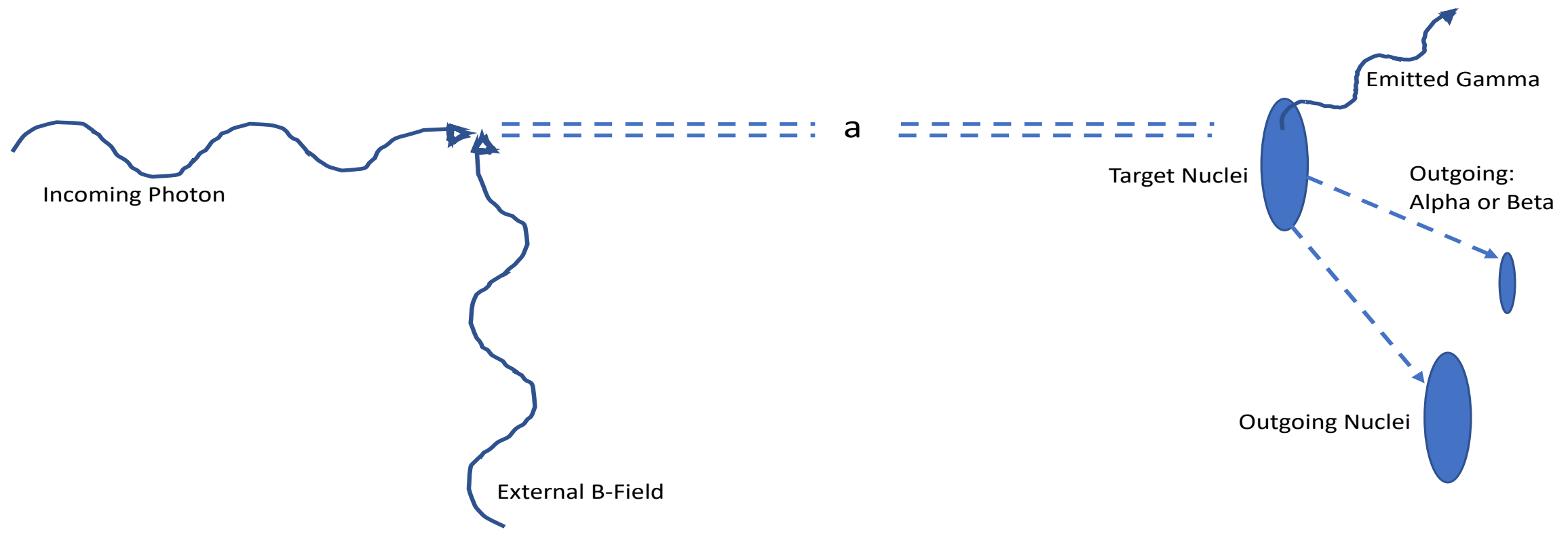

Figure 1: Primakoff production of axions (Left) and axion coupling to nuclear material (Right).

### $^{241}$Am:

$^{241}$Am results from a beta decay of plutonium ($^{241}$Pu). The isotope has a half-life of about 432 years. Notably it is one of the few isotopes that finds use in industry, in homes across the country to detect smoke. $^{241}$Am undergoes alpha decay to produce $^{237}$Np in an excited state. The excited $^{237}$Np emits intense gamma lines with energies of 59.54 and 26.32 keV. These lines are often used to calibrate nuclear detectors such as the NaI detectors used in this experiment.

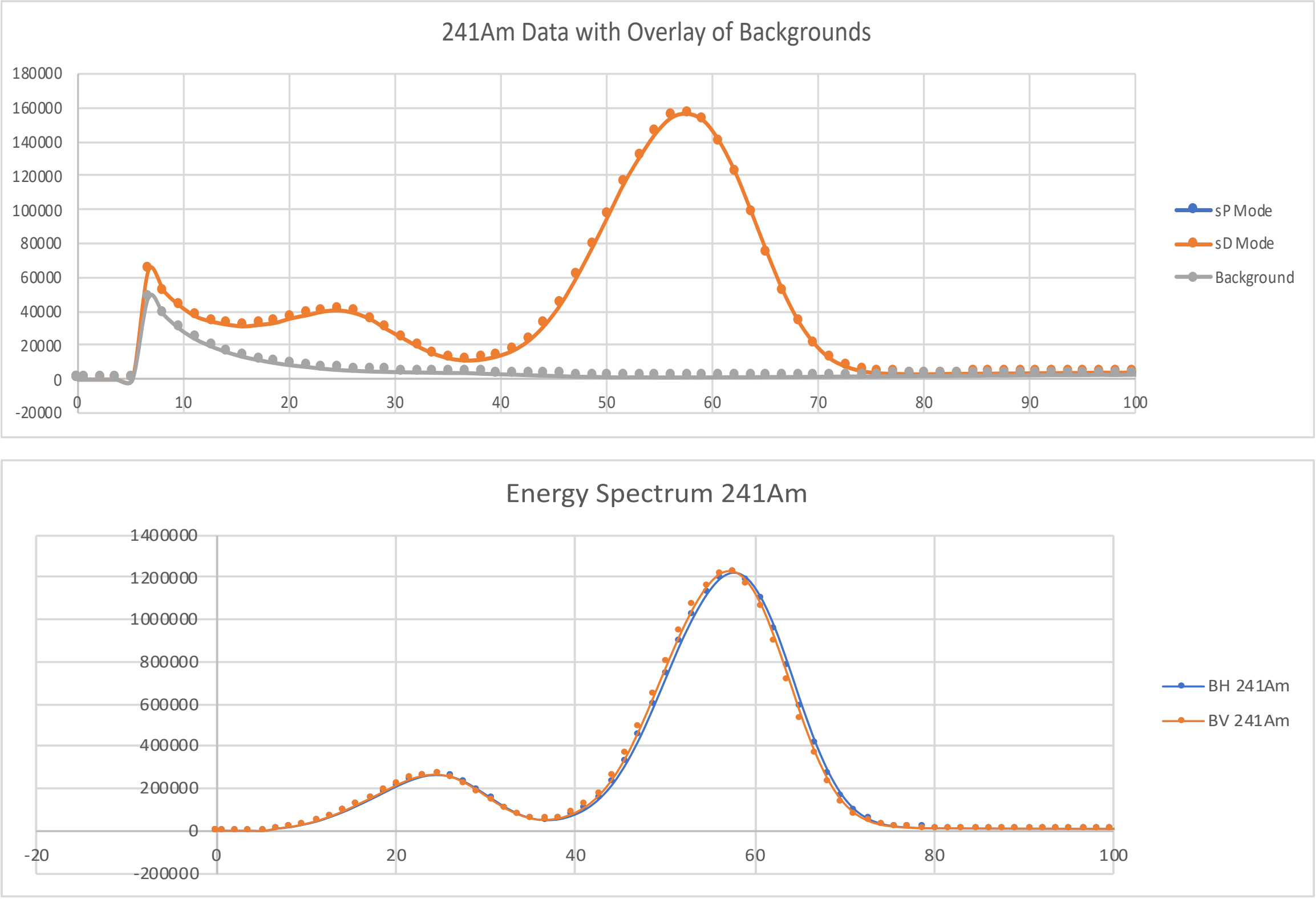

Figure 2: $^{241}$Am decay energy spectrum showing before (top panel) and after (bottom panel) background (top panel in gray) subtraction, collected using NaI gamma detectors.

Figure 2 shows a sample of data from the detection of $^{241}$Am decays. The plots references “sP” and “sD” modes to distinguish the experimental conditions described more below. At the resolution of this plot, there is no detectable difference between these modes. The analysis shown below reveals differences at a scale of ~ 0.33% or ~ 10 thousand out of 3 Million. Additionally, the bottom panel shows what happens when an external field is applied in either the Horizontal direction (BH) or Vertical direction (BV) relative to the optical bench. Rotation of the field between modes was found to cause a splitting of the measured energy spectrum as discussed in this paper.

**Experimental Setup:**

To detect any weakly interacting particles produced via the Primakoff mechanism, we propagate a beam of photons through a cavity which uses neodymium magnets to create a strong magnetic field. At the exit of the magnetic cavity, barriers prevent the light from directly interacting with the any of the radionuclides placed upstream where any weak particles should flow if produced in this process. The data is subdivided into an optical mode where the beam is allowed to pass through the cavity ("sP" indicated throughout this paper) and a dark mode where the beam is blocked before entering the cavity ("sD" indicating throughout this paper). The laser beams are always on to prevent any electronic effects associated with changing power drawn by turning on-off the beam.

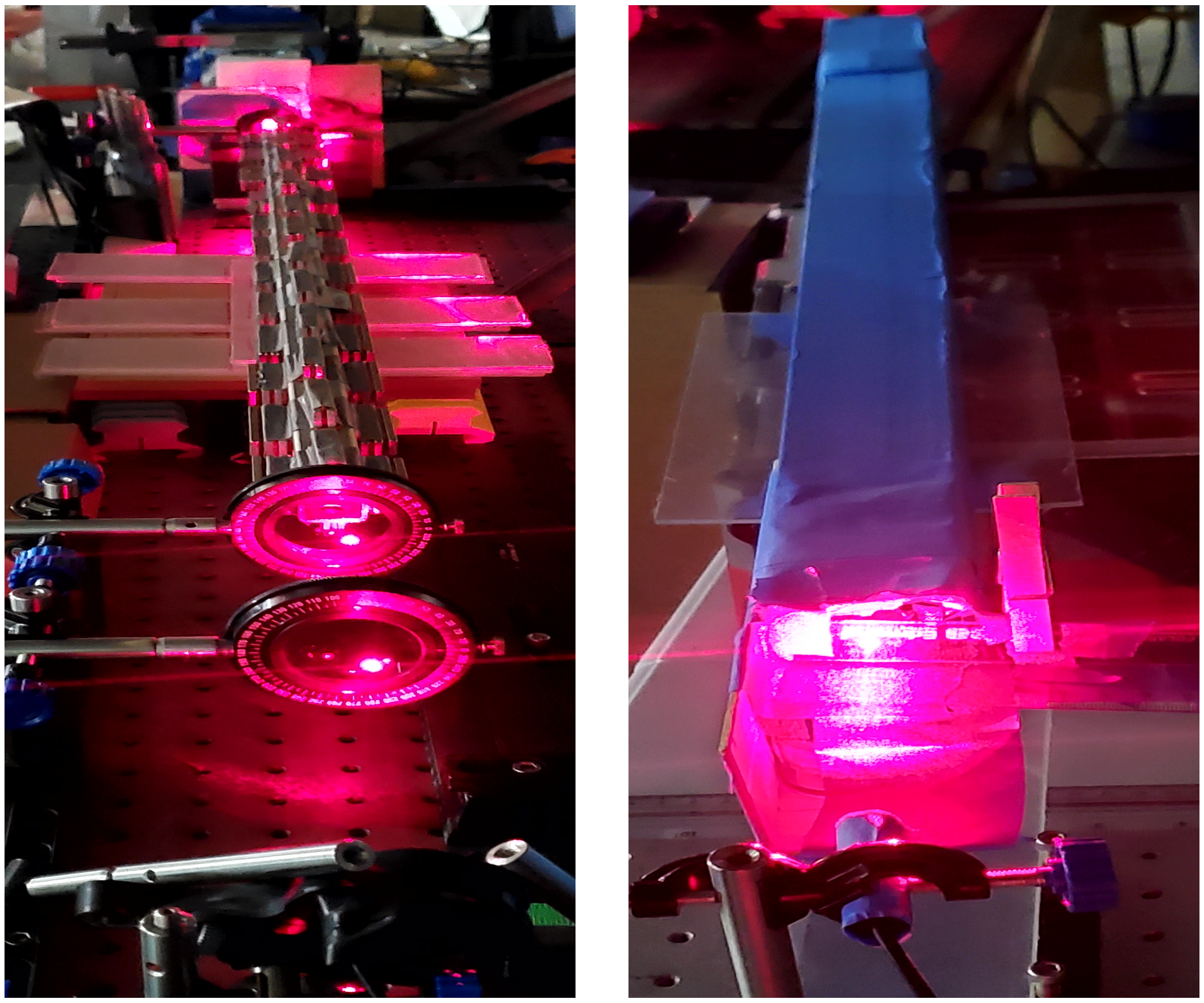

Figure 3: Magnetic cavity showing beams propagating through the magnetic fields.

Figure 3 shows two of the magnetic field apparatus used in taking the data analyzed in this work. The data is taken in an optical mode (sP) for 1 hour followed by a dark mode (sD) for 1 hour and then oscillated thereafter. A second detector is used to monitor the background radiation levels throughout the process. Because the second detector has a different, natural calibration relative to the detector used for the experimental data collection, the experimental detector is run in a "No Source" mode to collect both sD and sP data for background subtraction.  In this way, the second detector is only used to establish the relative background levels, that are then used to set the amplitude for the No Source data.  It is the No Source data taken on the experimental detector that is subtracted to yield the $^{241}$Am energy spectrum to calculate the final results.

The magnetic field is position to be either aligned in the plane of the optical bench and perpendicular to the propagation direction of the beam, this is "BH" or B-field horizontal or the field is aligned perpendicular (vertical) to both the plane of the optical bench and the direction of propagation for the optical beam "BV" mode.

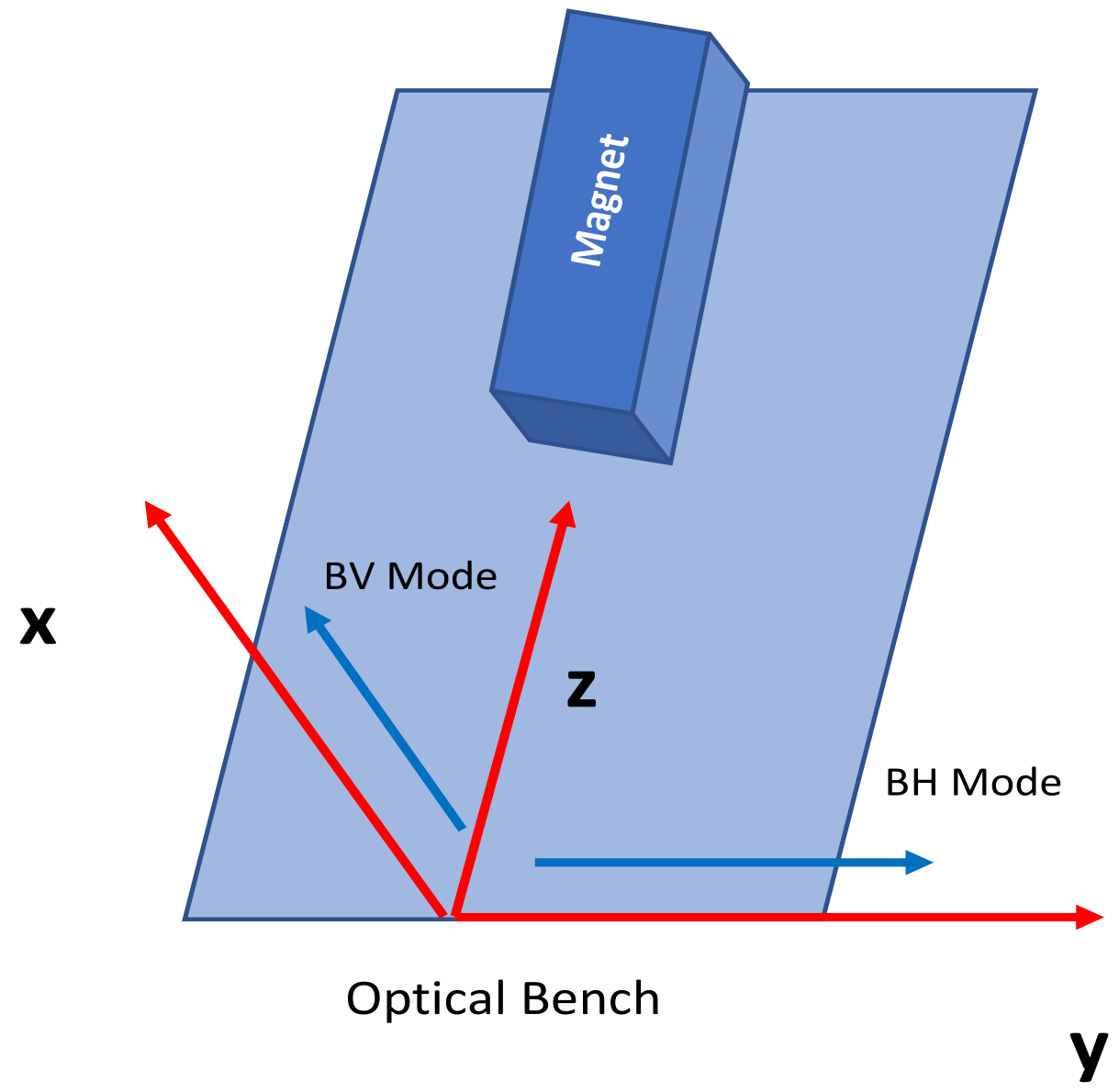

Figure 4: Experimental setup showing the magnetic cavity and direction (z) for beam propagation along with the two orientations for the magnetic field within the cavity (BH and BV) which are referenced throughout this paper.

The laser beam propagates along the z-direction and is polarized along the y-direction. The magnet is rotated such that it is aligned with the laser polarization ("BH") or perpendicular to the laser polarization ("BV").  The magnet is rotated only once to switch modes and is in the BH or BV configuration throughout data taking for any given set of experiments.

Table 1 in the Appendix of this paper gives the experimental parameters for the current setup.  These parameters can be used to calculate a final cross-section for the observed effects. They are also important to reproduction/confirmation of the exact measurements.  As the measurement involves a product of two cross-sections, calculating values for $g_{\alpha\gamma\gamma}$ and $g_{\alpha n}$ requires further measurements, the product of the associated cross sections are in Appendix 3.

**Data Analysis:**

As a function of the direction of the direction of the external field, the gamma spectrum from $^{237}$Np excited states was measured.

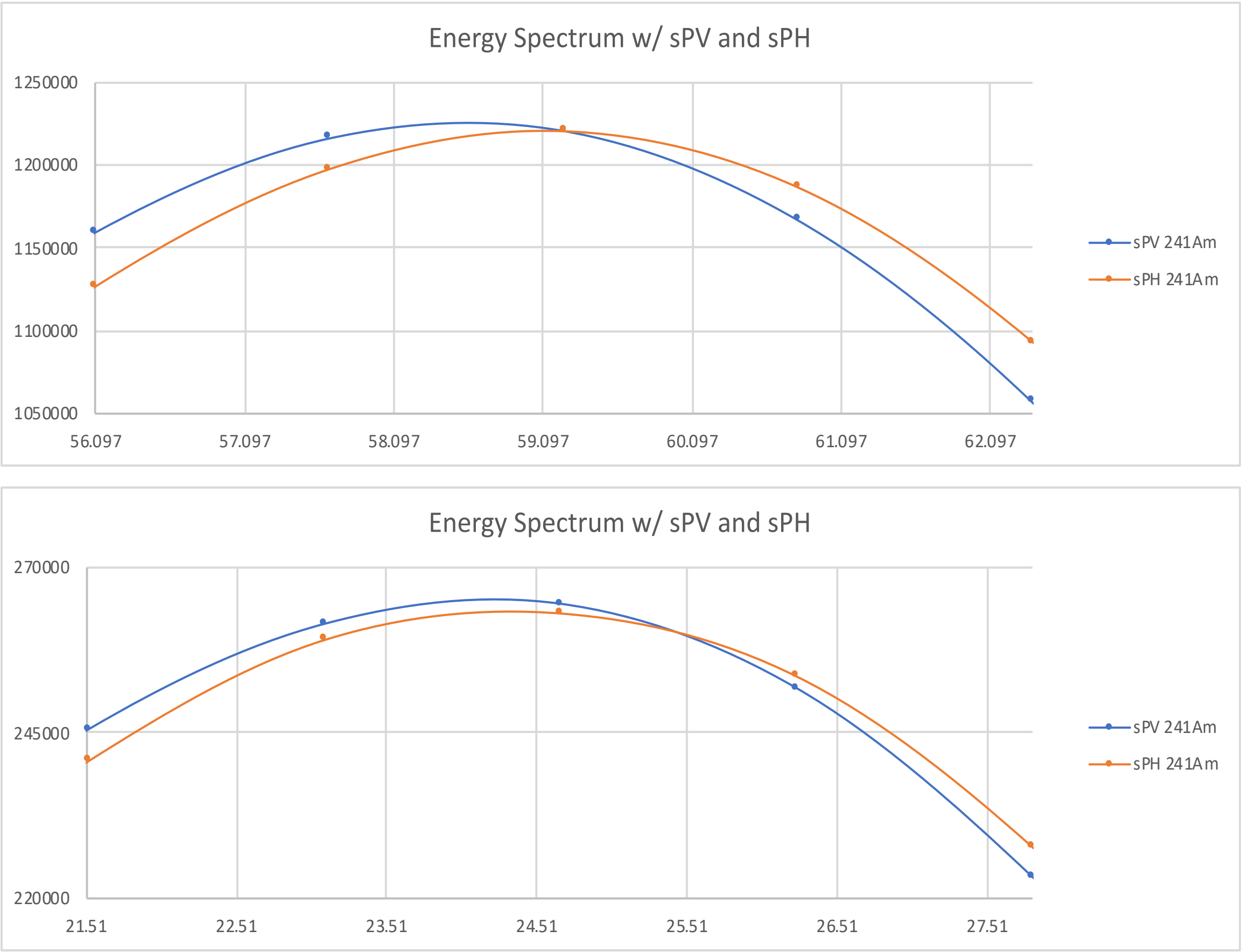

Figure 5: Data showing Integrated Counts over 18hours for Dark (sD) and Light (sP) modes for BH setup with arbitrarily scaled values for the statistical significance overlaid (top panel), sD – sP Raw Counts (middle panel) and the statistical significance, calculated by using a sliding window of width 12-bins equal to width of 59.54 gamma peak, and requiring that $\sigma > 4.5$ (bottom panel).

The data from $^{237}$Np showed a statistically significant difference for the two directions of applied, external magnetic field labeled: B Field Horizontal (BH) and B Field Vertical (BV). By contrast, the backgrounds shown in Figure 6, consisting mostly of gammas from daughter particle for $^{238}U$ such as $^{222}Rn$, showed no significant shifting in energy spectrum. After 18 hours, a 6.49$\sigma$ result was observed at approximately the center of the 59.54 keV peak. This suggests that the $^{241}Am$ is responding to some difference between the BH and BV.

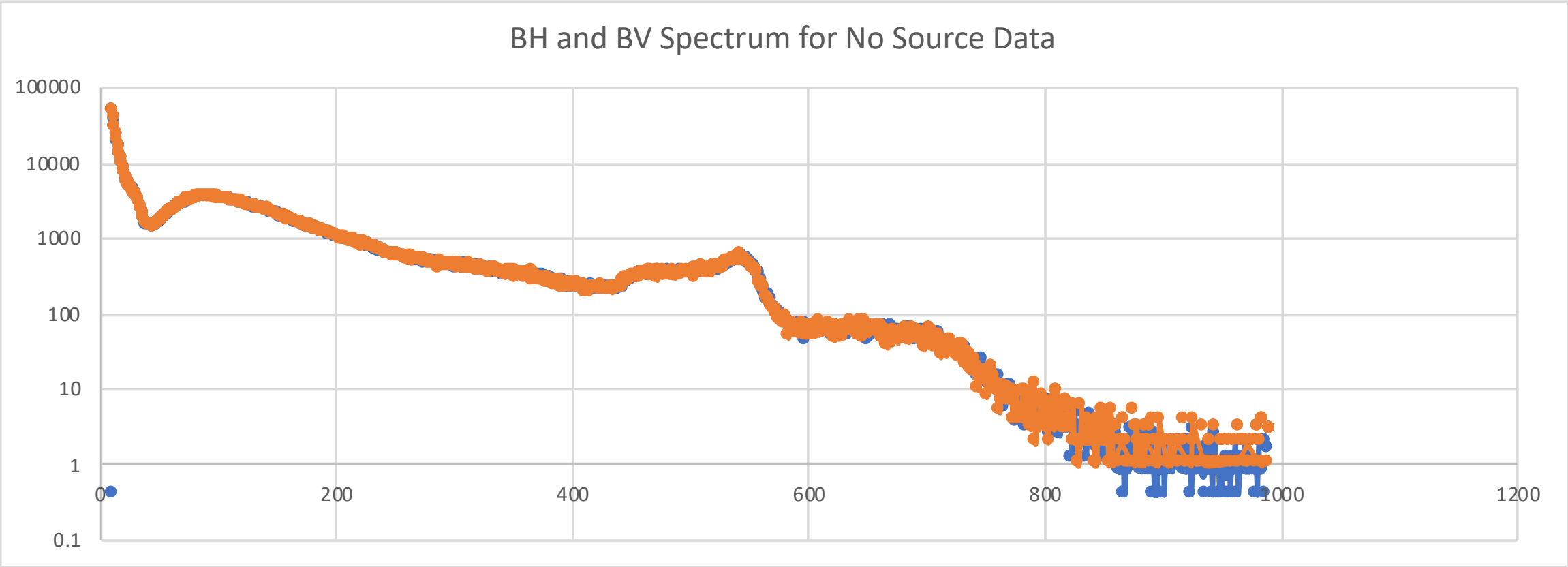

(a)

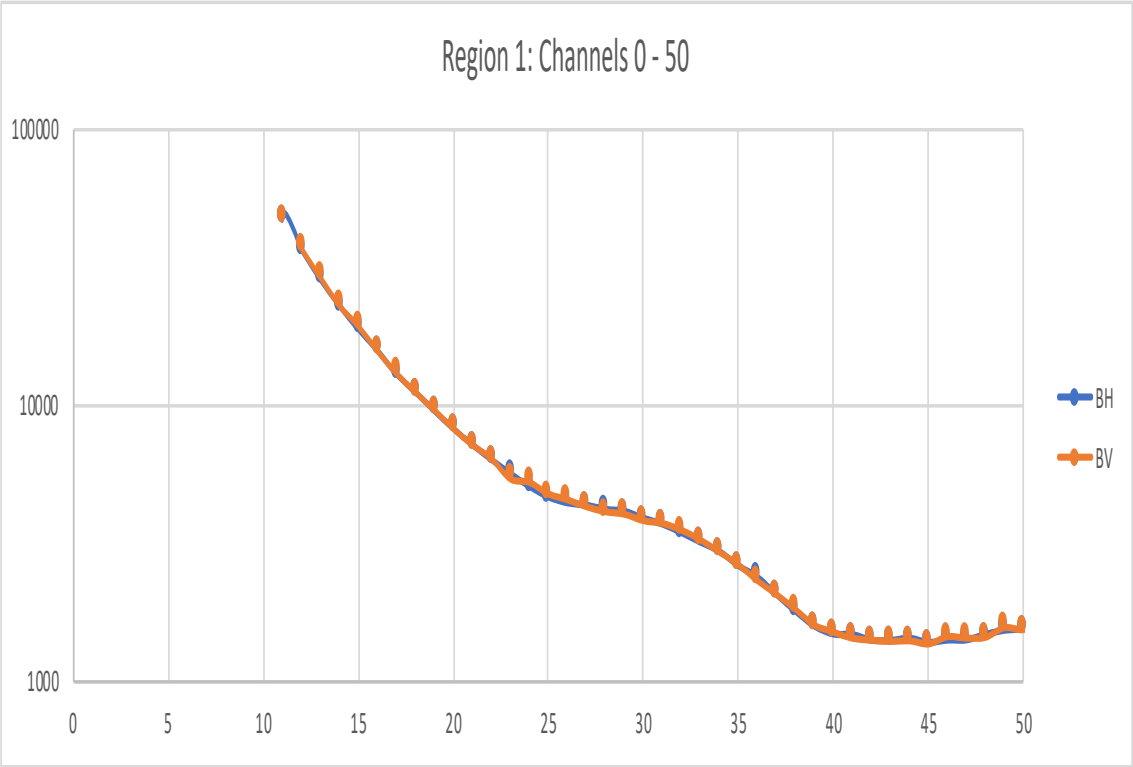

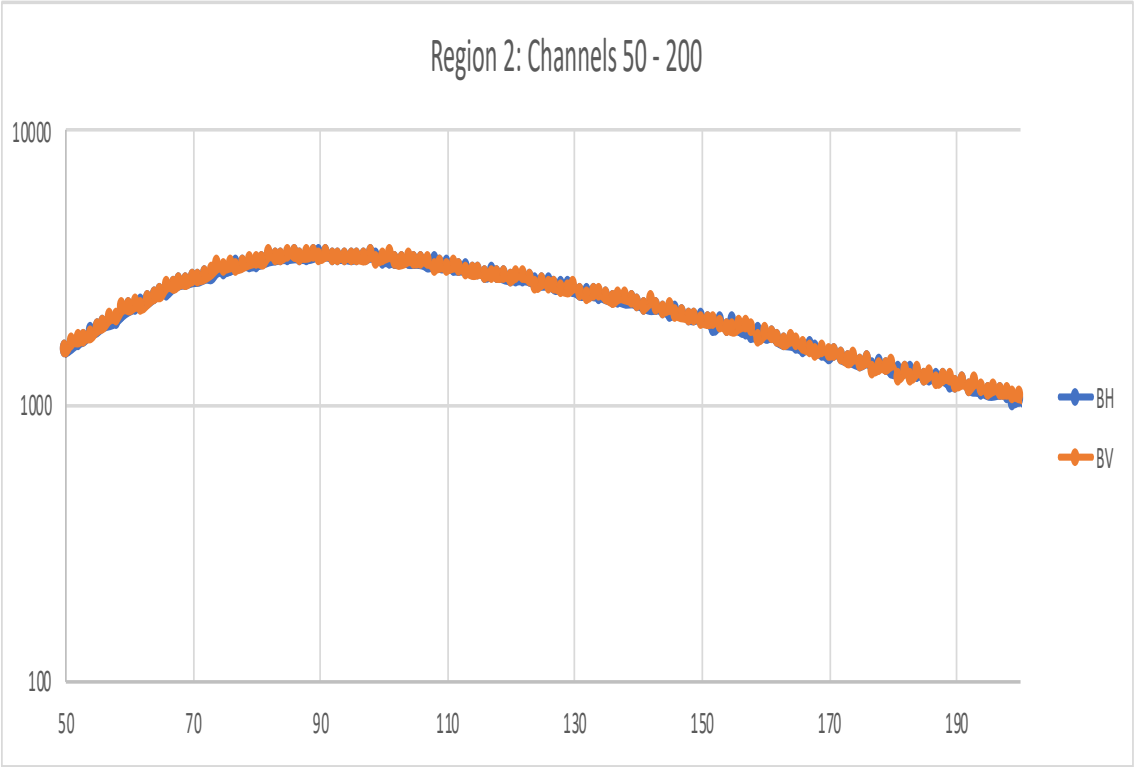

(b) (c)

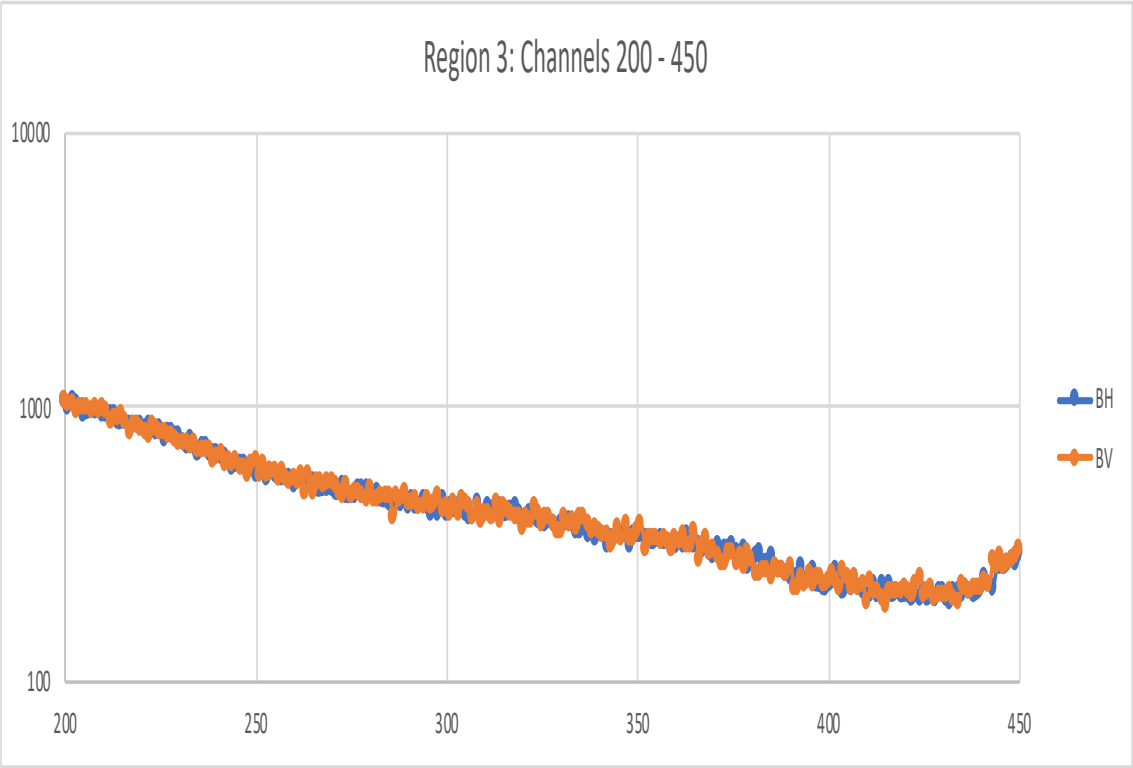

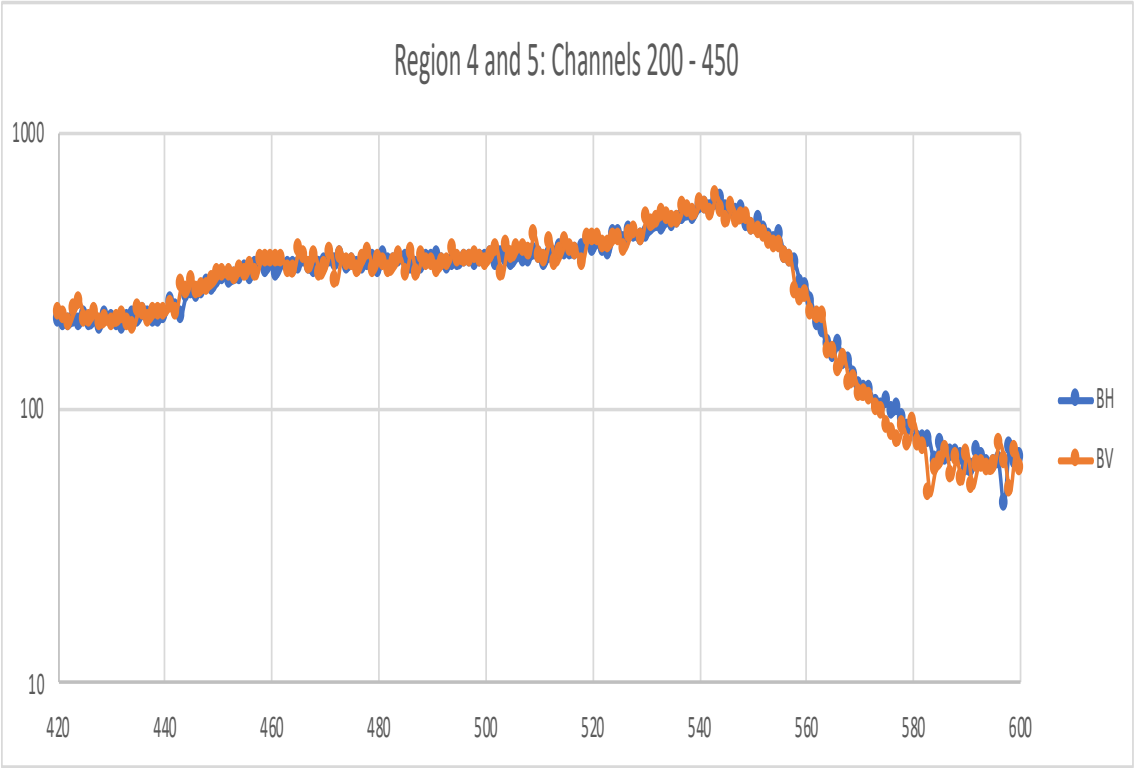

(d) (e)

Figure 6: Data showing full spectrum for background Radon events with B Field Horizontal (BH) and Vertical (BV) overlaid (a) along with several regions of interest (b) – (e).

In theory, the alignment of the nuclear magnetic dipole with an external field can be detected as a shifting in the energy states of the isotope. Table 1 below shows that for the no-source, background data, the regions corresponding to features from background radon do not show an energy shift as seen

**Table:** Analysis of the count rates for BV and BH for background only (Figure 6)

| | | | | | | |
|---|---|---|---|---|---|---|
| **Region 1** | | Mag_Shift | BV - BH | 475.4 | 188.4 | **0.43235914** |
| | | | | | | |
| **Region 2** | | Mag_Shift | | 831.8 | 696.4 | **0.08860097** |
| | | | | | | |
| **Region 3** | | Mag_Shift | | 262.8 | -142.6 | **0.26527941** |
| | | | | | | |
| **Region 4/5** | | Mag_Shift | | 537.2 | -138 | **1.69138277** |
| | | | | | | |
| **Region 4** | | Mag_Shift | | 269.6 | 34.6 | **0.77251808** |
| | | | | | | |
| **Region 5** | | Mag_Shift | | 225.6 | -368.6 | **-4.1552448** |
| | | | | | | |
| **Region 6** | | Mag_Shift | | -52.6 | -66.6 | **-0.1174497** |

**Conclusion:**

The current data set shows a statistical significant shifting of the 241Am spectrum depending on the direction of the externally applied magnetic field. Though the field at the 241Am source is only about 5G, the degree of shifting creates a statistically significant effect.